\documentclass[aps,rmp,preprint,groupedaddress]{revtex4-1}
\usepackage{setspace}
\begin{document}

%\preprint{}

\title{QCD beyond diagrams}


\author{Michael Creutz}
\email{mike@latticeguy.net}
%\thanks{}
\affiliation{Brookhaven National Laboratory
\\ Senior Physicist Emeritus}


%\date{9 october 2020}

\begin{abstract}
  QCD, the theory of the strong interactions, involves quarks
  interacting with non-Abelian gluon fields.  This theory has many
  features that are difficult to impossible to see in conventional
  diagrammatic perturbation theory.  This includes quark confinement,
  mass generation, and chiral symmetry breaking. This paper is a
  colloquium level overview of the framework for understanding how
  these effects come about.
\end{abstract}

%\keywords{}

\maketitle

\section{Introduction}

\input epsf

% to comment out parts of a TeX file
\long \def \blockcomment #1\endcomment{}

%%%%%%%%%%%%%%%%%%%%%%%%%%%%%%%%

The standard model of particle physics combines electrodynamics, the
weak interactions responsible for beta decay, and the strong
interactions of nuclear physics.  The strong interaction sector is
based on quarks interacting via the exchange of non-Abelian gluon
fields, and is referred to as ``QCD'' or ``quantum chromodynamics.''
It is in some sense the best defined part of the standard model.
Unlike electrodynamics, the behavior of QCD is controlled at short
distances by the phenomenon of ``asymptotic freedom.''  While
electrodynamics may have numerous successes, it remains unclear if it
is well defined on its own.  The weak interactions have additional
issues associated with the absence of a rigorous non-perturbative
scheme for incorporating parity violation.

Most of our quantitative understanding of quantum field theory comes
by way of perturbative expansions.  However, QCD has several features
that are difficult to impossible to understand in terms of Feynman
diagrams.  First is confinement; quarks do not propagate as free
particles, but always appear in color singlet bound states.  Second,
the theory generates its own mass scale; even if the quarks are
massless, the proton is not.  Third is the presence of chiral symmetry
breaking, responsible for the pion being much lighter than the rho,
although both are made of the same quarks.  And finally, the theory
depends on a hidden parameter, usually called ``Theta.''  Different
values of Theta represent physically inequivalent theories, despite
having identical perturbative expansions.  The framework for
investigating how these phenomena come about in QCD is the subject of
this paper.  For those interested in more details, see the book
\underline {From Quarks to Pions}
\cite{Creutz:2018dgh}.

\section{Path integrals}

\begin{figure}[h] \em
\epsfxsize .4\hsize
\centerline{\epsffile{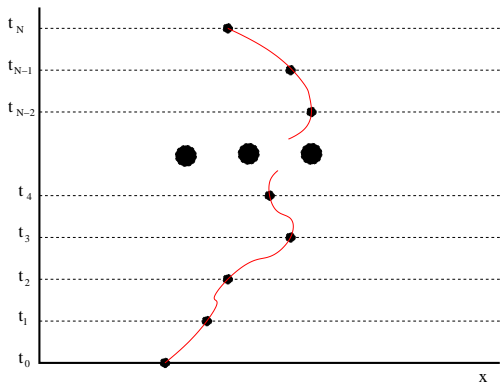}}
 \caption{\label{integral}
 Dividing time into slices to define the path integral.}
 \end{figure}

Our discussion revolves around the Feynman path integral formulation
of quantum mechanics.  Feynman's classic 1948 article demonstrated a
remarkable equivalence between a quantum mechanical system and a
statistical mechanics problem in one more dimension \cite
{Feynman:1948ur}.  Consider a particle moving in some potential $V(x)$
and an action function for an arbitrary path $x(t)$ between $t=0$ and
$t=\beta$
\begin{equation}
  S=\int_0^\beta  dt\ \dot x^2/2+V(x)
\end {equation}
Feynman showed that by integrating over all possible paths, there is
an equivalence to the quantum mechanics of a unit mass particle moving
in the potential $V$ with Hamiltonian
\begin {equation}
  H= \hat p^2/2+V(\hat x)
\end{equation}
where $\hat x$ and $\hat p$ are quantum mechanical operators
satisfying the canonical commutation relation
\begin{equation}
[\hat
  p,\hat x]\equiv \hat p \hat x - \hat x \hat p=i.
\end{equation}
Formally we have
\begin{equation}
\int (dx(t))\ e^{- S(x(t))} \qquad \propto \qquad {\rm
  Tr}\ e^{-\beta H}
\end{equation}
where the trace is over the quantum mechanical Hilbert space.  To
define what is meant by integrating over paths, Feynman introduced a
discretization of time in steps of size $a$ and then integrated over
the position on each time slice, as sketched in Fig. \ref{integral}.
The path integral is defined in a limit $a\rightarrow 0$.

This process proceeds similarly for field theories, with a $D$
dimensional quantum field theory related to a $D+1$ dimensional
statistical mechanics problem.  In field theory the ``paths'' are
usually referred to as ``configurations.''  This approach is the heart
of the Monte Carlo approach to lattice gauge theory.

In this picture, a path looks much like a wiggling string or
worm.  As the time spacing becomes small, this worm becomes
increasingly wiggly.  Indeed, in the limit $a\rightarrow 0$ the
typical paths are not differentiable in the sense that
\begin{equation}
\langle \dot x ^2\rangle
= \left\langle \left ( {x_{i+1}-x_i\over
a} \right)^2\right \rangle
\propto {1\over a}\rightarrow \infty.
\end{equation}
This roughness extends to the field theory case, wherein the typical
configurations are also non-differentiable.  This raises interesting
ambiguities to which we return in Section \ref{masses}.

\section{Perturbation theory} 

At this point traditional treatments of quantum field theory turn to
perturbation theory.  Divide the action into two parts
\begin{equation}
  S=S_0+gS_I
\end{equation}
where $S_0$ describes a solvable theory of free particles and $S_I$ couples
the fields together.  Expanding the path integral in the coupling $g$
gives the usual Feynman diagrams wherein free propagators are coupled
by interaction vertices.  But, for our purposes, it is important
to realize that the path integral approach is much more general.

\begin{figure}[h] \em
\epsfxsize .5\hsize
\centerline{\epsffile{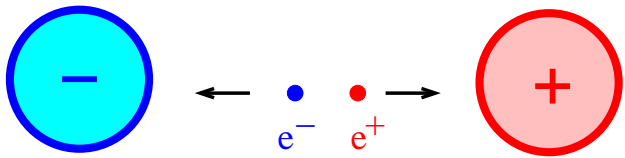}}
\caption{\label{perpetual}
    If like charges attracted, one could
   create an unstable situation by creating two large regions of
   opposite charge and tossing particle-antiparticle pairs into them.}

  \end{figure}

For the electro-weak part of the standard model, the basic coupling
$\alpha={g^2\over 4 \pi}\sim 1/137$ is a small number, and
perturbation theory works well for all practical purposes.  However
Dyson provided a simple argument that perturbation theory cannot
converge \cite{Dyson:1952tj}.  He argued that if it did, then one
could analytically continue $e\rightarrow ie$ and we would have a
theory where like charges attract instead of repel.  In this case one
could produce separate regions of charge large enough to bind
additional electrons by more than their rest mass.  One could use this
situation to create a perpetual motion machine by creating electron
positron pairs and sending them to the respective charged regions.
This is sketched in Fig.  \ref{perpetual}. The vacuum of such a theory
is unstable.\footnote{The reader might note a similarity between this
  argument and the presence of Hawking radiation from a black hole.
  The main difference is that Hawking radiation reduces the mass of
  the black hole, while the process described here increases the net
  charge of the corresponding orbs.}
  
  \begin{singlespace}
\end{singlespace}

\begin{figure}[h] \em
\epsfxsize .4\hsize
\centerline{\epsffile{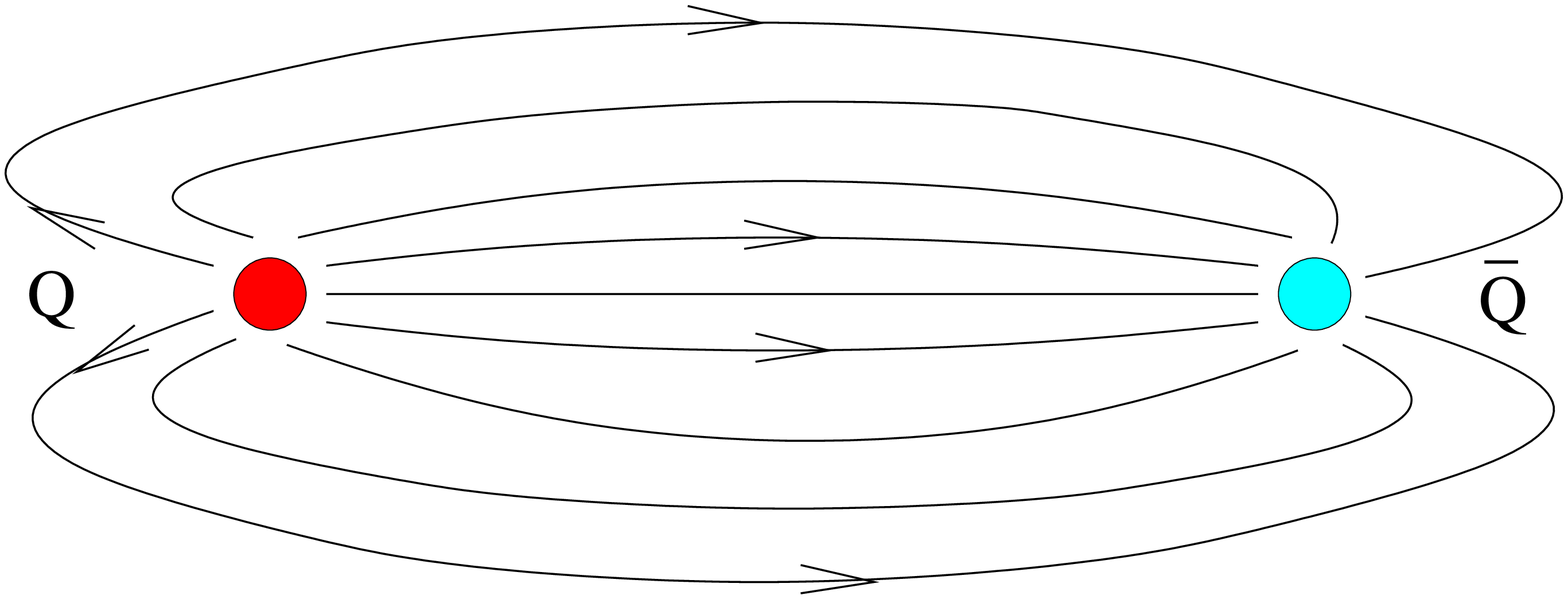}}
\caption{\label{fluxtube} The gluon electric field between quark pairs
  becomes concentrated into a tube of flux with a net energy per unit
  length.  The tension in this string is roughly 14 tons.}
\end{figure}

The problems with perturbation theory are particularly acute with QCD,
the theory of quarks and gluons.  If the coupling vanishes, $g=0$, the
physical propagating fields are quarks and massless gluons.  However
as soon as $g \ne 0$ we have a theory where the physical particles
are protons and pions.  This is a qualitatively different spectrum.
The gluon electric fields between quark charges change from a
Coulombic spread into a flux tube giving a linear potential, as
sketched in Fig. \ref{fluxtube}.  To understand these qualitative
changes, one must go beyond perturbation theory.

\section{Divergences and renormalization}

As is well known, relativistic quantum field theory is replete with
divergences that require renormalization.  One must introduce a
cutoff, and then remove it by a limiting procedure.  The basic idea is
to fix a few physical observables and adjust the bare couplings and
masses in such a way that the physics remains finite as the cutoff
is removed.  In the process the bare charges and masses absorb the
divergences by either going to infinity or zero.

\begin{figure}[h]
  \em
\epsfxsize .3\hsize
\centerline{\epsffile{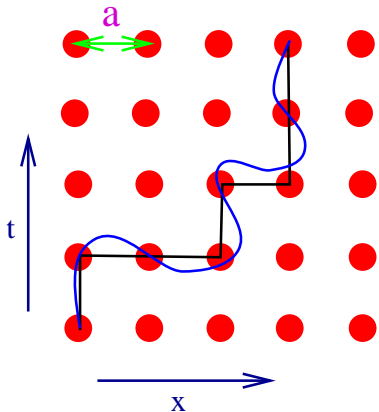}}
\caption{\label{worldline} On the lattice a world line of any particle
  is replaced by a sequence of discrete hops. The size of these hops
  represents an ultraviolet cutoff and should be taken to zero for the
  continuum limit.}
\end{figure}

Historically most regulators used in practice, such as Pauli-Villars
or dimensional, are based on perturbation theory.  One starts to
calculate and on finding a divergent diagram inserts the cutoff.  But
QCD requires a non-perturbative regulator.  This is where lattice
gauge theory comes in, with the lattice spacing $a$ being a cutoff.
In terms of the world lines a quark might take, they are replaced by
single steps on the lattice, as sketched in Fig. \ref{worldline}.
Most lattice formulations are based on Wilson's \cite{Wilson:1974sk}
elegant picture of quarks hopping between sites while picking up
non-Abelian phases on the bonds.  We return to this in Section
\ref{thelattice}.

As a simple example of how renormalization might theoretically
proceed, consider taking the proton mass as the physical
observable.\footnote{Because of difficulties calculating with light
  quarks, usually one uses other physical quantities to hold fixed.
  One example is the mass of the $\Omega^-$ particle since massive
  strange quarks are easier to handle \cite{Capitani:2011fg}.} This is
expected to remain finite, even if the quark masses vanish.  Ignoring
quark masses for the moment, the proton mass is a function of the
lattice spacing $a$ and the coupling $g$.  We want to vary the
coupling for the continuum limit so the proton mass is constant.  For
this we have
\begin{equation}
a{d\over da}m_p(a,g)=0=
\left( {\partial \over \partial g} m_p \right)
\left( a{d g\over da}\right)+
a{\partial\over \partial a}m_p.
\end{equation}
Since the only dimensional input is the lattice spacing, dimensional
analysis tells us
\begin{equation}
a{\partial\over \partial a}m_p=-m_p.
\end{equation}
Putting this together, we conclude
\begin{equation}
a{d g\over da}\equiv \beta(g)
={m_p \over {\partial\over \partial g}
  m_p}.
 \label{rgeq}
\end{equation}
This totally non-perturbative definition shows formally how $g$
should depend on $a$ for the physical limit.

\section{Asymptotic freedom}

While shirking perturbation theory to this point, a remarkable feature
of the function $\beta(g)$ is that it does itself have an
expansion in $g$, and this expansion gives non-perturbative
information on how to take the continuum limit. 
Gross, Wilczek, and Politzer in their Nobel Prize winning papers showed
\cite{Gross:1973ju,Politzer:1973fx}
\begin{equation}
  \label{af}
  \beta(g)=\beta_0 g^3 + \beta_1 g^5 + O(g^7)
\end{equation}
where
\begin{equation}
  \beta_0={1\over 16\pi^2}(11-2N_f/3)
 \end{equation}
and $N_f$ is the number of distinct quark ``flavors'' under
consideration \cite{Gross:1973ju,Politzer:1973fx}.
%\footnote{I use a
%  different sign convention for $\beta$ from that in the quoted
%papers.}
 A year later Caswell and Jones\cite{Caswell:1974gg,Jones:1974mm}
 calculated
\begin{equation}
  \beta_1=\left({1\over 16\pi^2}\right)^2 (102-22N_f/3).
\end{equation}
A remarkable feature of this expansion is that $\beta_0$ and $\beta_1$
are independent of the details of the regularization scheme, although
higher terms are not.  This means that this result is immediately
applicable to the lattice as well.

The differential equation obtained by combining Eqs. (\ref{rgeq})
and (\ref{af})
\begin{equation}
a{d g\over da}=\beta_0 g^3 + \beta_1 g^5 + O(g^7)
\end{equation}
is easy to solve, giving
\begin{equation}
  \label{avsg}
 a={1\over \Lambda} \ 
g^{-\beta_1/\beta_0^2}\ \exp\left({-1\over 2\beta_0 g^2}\right)
\ (1+O(g^2))
\end{equation}
where $\Lambda$ is an integration constant.  Note the factor of
$\exp\left({-1\over 2\beta_0 g^2}\right)$ has an essential singularity
and cannot be expanded in a power series in $g$.  This equation also
shows that taking the lattice spacing to zero requires $g$ to go to
zero.  This is what is called ``asymptotic freedom.''

The integration constant $\Lambda$ has dimensions of mass.\footnote{In
  units where both $\hbar$ and $c$ are unity.}  While the
classical theory with massless quarks contains no dimensional scale,
renormalization forces us to bring one in.  All particle masses,
including the proton, are proportional to $\Lambda$
\begin{equation}
 m_p\propto \Lambda \propto {1\over a}\
g^{-\beta_1/\beta_0^2}\ 
\exp\left({-1\over 2\beta_0 g^2}\right)
\end{equation}
again showing non-perturbative behavior!  The idea that
renormalization can eliminate a dimensionless coupling $g$ while
generating an overall mass scale has been given the marvelous name
``dimensional transmutation'' \cite{Coleman:1973jx}.

\section {Mass renormalization}

\begin{figure}[h] \em
\epsfxsize .35\hsize
\centerline{\epsffile{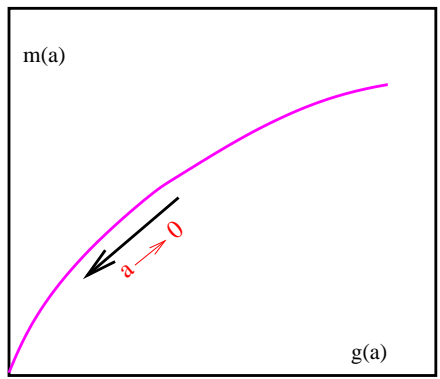}}
\caption{\label{rgflow} For the continuum limit, both the bare coupling and
  quark masses flow towards vanishing values.}
\end{figure}

So far we have ignored the masses of the quarks.  These also need to be
renormalized, requiring additional physical things to be fixed for the
continuum limit.  That brings in more integration constants.
We'll be brief here and only note that the bare mass $m$ of any
quark flavor should satisfy
\begin{equation}
  a{dm\over da}=m\gamma(g)=m(\gamma_0 g^2+\dots)+\hbox{ non
    perturbative}
  \label{mrg}
\end{equation}
where $ \gamma_0={8\over (4\pi)^2 } $ \cite{Georgi:1976ve}. The
``non-perturbative'' expression in the above equation allows terms
that fall faster than any power of $g$ and will play a role in later
discussion.  As before, this equation is easily integrated
\begin{equation}
  m=M\ g^{\gamma_0/\beta_0}\ (1+O(g^2))
\end{equation}
where $M$ is another integration constant.  There is one such $M_i$
corresponding to each quark species $i$.  Note that since $g$ goes to
zero for the continuum limit, so do the bare quark masses $m_i$.  The
full continuum requires taking both the bare $g$ and $m$ together, as
sketched in Fig \ref{rgflow}.

The continuum theory depends only on the dimensional integration
constants $M_i$ and $\Lambda$.  Once these are known, all physical
results are in principle determined.  While they depend on the
physical quantities held fixed for renormalization, their specific
values depend on the details of the regulator.  As will be discussed
in Section \ref{masses}, they can mix together depending on the
details of the scheme.  The physical observables being held fixed,
such as the proton and pseudoscalar masses, are all that the theory
ultimately depends on.

\section{The lattice}
\label{thelattice}

The primary role of lattice gauge theory is as a non-perturbative
regulator.  It provides a minimum wavelength through the lattice
spacing $a$, {\it i.e.} a maximum momentum of $\pi/a$.  Avoiding the
convergence difficulties of perturbation theory, the lattice provides
a route towards a rigorous definition of a quantum field theory as a
limiting process.  

As formulated by Wilson, the lattice cutoff remains true to many of
the underlying concepts of a gauge theory.  In particular a gauge
theory is a theory with a local symmetry.  With the Wilson action
being formulated in terms of products of group elements around closed
loops, this symmetry remains exact even with the regulator in place.

The concept of gauge fields representing path dependent phases leads
directly to the conventional lattice formulation.  We approximate a
general quark world line by a set of hoppings lying along lattice
bonds, as sketched earlier in Fig. \ref {worldline}.  The gauge field
corresponds to $SU(3)$ phases acquired along these hoppings.  Thus the
gauge fields are a set, $\{U\}$, of group matrices, one element
associated with each nearest neighbor bond of the four-dimensional
space-time lattice.

\begin{figure}[h]
  \em
\epsfxsize .25\hsize
  \centerline{
\epsffile{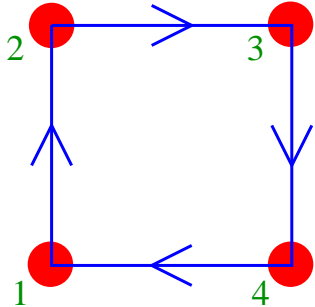}
}
\caption{Analogous to Stoke's law, the flux through an elementary
  square of the lattice is found from the product of gauge matrices
  around that square.  The dynamics is determined by adding the real
  part of the trace of this product over all elementary squares.
\label{plaquette}
}
\end{figure}

Motivated by the electromagnetic flux being the generalized curl of
the vector potential, we identify the flux through an elementary
square, or ``plaquette,'' on the lattice with the phase factor
obtained on running around that plaquette; see Fig. \ref {plaquette}.
Spatial plaquettes represent the ``magnetic'' effects and plaquettes
with one time-like side give the ``electric'' fields.  This
motivates the conventional ``action'' used for the gauge fields as a
sum over all elementary squares of the lattice.  Around each square we
multiply the phases and take the real part of the trace
\begin{equation}
S_g=\sum_p {\rm Re\ Tr} \prod_{l\in p} U_l \sim \int d^4x\ E^2+B^2.
\end{equation}  
Here the fundamental squares are denoted $p$ and the links $l$.  As we
are dealing with non-commuting matrices, the product around the square
is meant to be ordered, while, because of the trace, the starting point
of this ordering drops out.

For the quantum theory we turn to the path integral.  To construct
this, exponentiate the action and integrate over all dynamical
variables
\begin{equation}
Z=\int (dU) e^{-\beta S},
\end{equation}
where the parameter $\beta$ controls the bare coupling.  This converts
the three space dimensional quantum field theory of gluons into a
classical statistical mechanical system in four space-time dimensions.

The formulation is conventionally taken in Euclidean four-dimensional
space.  In effect this replaces the time evolution operator $e^{-iHt}$
by $e^{-Ht}$.  Then the path integral involves positive weights, and
can be studied by standard Monte Carlo methods.  This approach
currently dominates lattice gauge research.  The simulations have had
numerous successes, from demonstrating confinement
\cite{Creutz:1980zw} to reproducing the hadronic spectrum
  \cite{Durr:2010vn}.

  \section{Pions as pseudo-Goldstone bosons}

Much older a tool than the lattice, ideas based on chiral symmetry
have historically provided considerable insight into how the strong
interactions work.  In particular, chirality is crucial to our
understanding of why the pion is so much lighter than the rho, despite
being made of the same quarks. 

The classical Lagrangian for QCD couples left and right handed quark
fields only through mass terms.  If we project out the two helicity
states of a quark field by defining
\begin{equation}
\psi_L={1+\gamma_5\over 2}\psi\qquad\psi_R={1-\gamma_5\over 2}\psi
\end{equation}
then the classical kinetic term for the quarks separates
\begin{equation}
i\overline\psi D \psi=
i\overline\psi_L D \psi_L+
i\overline\psi_R D \psi_R.
\end{equation}
On the other hand, the mass term directly mixes the two chiralities
\begin{equation}
m\overline\psi \psi=
m\overline\psi_R \psi_L+
m\overline\psi_L \psi_R.
\end{equation}

Thus naively the massless theory appears to have independent conserved
quantities, one associated with each handedness.  After confinement,
it is generally understood that the axial chiral symmetry is
spontaneously broken via the scalar field $\sigma\propto \overline\psi
\psi$ acquiring a vacuum expectation value.  As with the mass term,
this field mixes chiralities.  This expectation value is not invariant
under the axial chiral symmetry.  Although chiral symmetry is broken,
parity is not; it is only the independence of the left and right quark
fields that the vacuum does not respect.  

For $N_f$ massless flavors, there is classically an independent
$U(N_f)$ symmetry associated with each chirality, giving what is often
written in terms of axial and vector fields as an $U(N_f)_V\otimes
U(N_f)_A$ symmetry.  As we will see in the following section, this
full symmetry does not survive quantization, being broken to a
$SU(N_f)_V\otimes SU(N_f)_A\otimes U(1)_B$, where the $U(1)_B$
represents the symmetry of baryon number conservation.  The only
surviving axial symmetries of the massless quantum theory are
non-singlet under flavor symmetry.

We work here with composite scalar and pseudo-scalar fields
\begin{equation}
\matrix{
\sigma&\sim&\overline\psi \psi\cr
\pi_\alpha&\sim& i\overline\psi \lambda_\alpha \gamma_5 \psi \cr
\eta^\prime &\sim& i\overline\psi\gamma_5\psi. \cr
}
\label{mesonfields}
\end{equation}
Here the $\lambda_\alpha$ are the generators for the flavor group
$SU(N_f)$.  They are generalizations of the usual Gell-Mann matrices
from $SU(3)$; however, now we are concerned with the flavor group, not
the internal symmetry group related to confinement.  As mentioned
earlier, we must assume that some sort of regulator, perhaps a
lattice, is in place to define these products of fields at the same
point.  Indeed, most of the quantities mentioned in this section are
formally divergent, although we concentrate on those aspects that
survive the continuum limit.

\begin{figure}[h] \em
      \epsfxsize .5\hsize
  \centerline{\epsffile{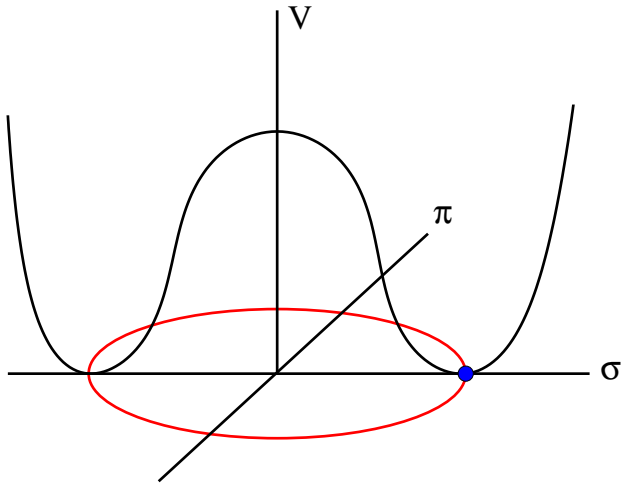}}
  \caption{\label{v0} The flavor non-singlet pseudo-scalar mesons are
  Goldstone bosons corresponding to flat directions in the effective
  potential.  This shape is sometimes referred to as a ``Mexican hat''
  or ``a wine-bottle bottom'' potential.  }
\end{figure}

\blockcomment
To simplify the discussion, consider degenerate quarks with a small
common mass $m$.    It is also convenient to initially
restrict $N_f$ to be even, saving for later some interesting
subtleties arising with an odd number of flavors.  And we assume $N_f$
is small enough to maintain asymptotic freedom as well as avoiding any
possible conformal phases.
\endcomment

The conventional picture of spontaneous chiral symmetry breaking in
the limit of massless quarks assumes that the vacuum acquires a quark
condensate with
\begin{equation}
\langle\overline\psi\psi\rangle
=\langle\sigma\rangle =v \ne 0.
\end{equation}
Extending the effective potential to a function of multiple
non-singlet pseudo-scalar fields gives the standard picture of
Goldstone bosons.  These are massless when the quark mass vanishes,
corresponding to $N_f^2-1$ ``flat'' directions for the potential.  One
such direction is sketched schematically in Fig.~\ref{v0}.  For the
two flavor case, these rotations represent a symmetry mixing the sigma
field with any of the pions
\begin{equation}
\matrix{
\sigma\rightarrow\phantom{+}  \sigma \cos(\phi)+\pi^\alpha \sin(\phi)\cr
\pi^\alpha\rightarrow -\sigma \sin(\phi)+\pi^\alpha \cos(\phi).\cr
}
\end{equation}
Pions are waves propagating through the non-vanishing sigma condensate
with oscillations in a direction ``transverse'' to the sigma
expectation.  They are massless because there is no restoring force in
that direction.

\begin{figure}[h] \em
\epsfxsize .5\hsize
\centerline{\epsffile{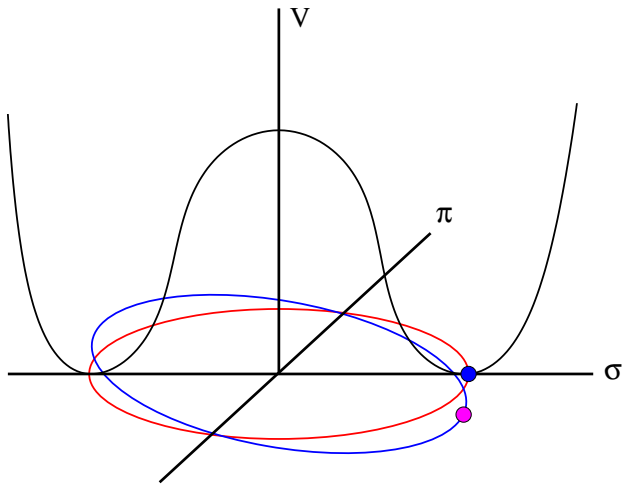}}
\caption{\label{v3}
A small quark mass term tilts the effective potential, selecting one
direction for the true vacuum and giving the Goldstone bosons a mass
proportional to the square root of the quark mass.
}
\end{figure}

If we now introduce a small mass for the quarks, this will effectively
tilt the potential $V(\sigma)\rightarrow V(\sigma)-m\sigma$.  This
selects one minimum as the true vacuum.  The tilting of the potential
breaks the global symmetry and gives the Goldstone bosons a small mass
proportional to the square root of the quark mass, as sketched in
Fig.~\ref{v3}.   With non-degenerate quarks, it is the average quark mass that gives the pseudoscalar mass; i.e. for the charged pion we have
\begin{equation}
  M_{\pi^+}^2 \propto {m_u+m_d \over 2}\Lambda.
  \label{averageqm}
\end{equation}
The standard chiral Lagrangian approach is a simultaneous expansion in
the masses and momenta of these light particles.

\blockcomment
In general, Goldstone bosons are associated with conserved currents
where the charges do not leave the vacuum invariant.  In the present
case these are the axial currents formally given by the quark
bilinears
\begin{equation}
A_\mu^\alpha=\overline\psi \lambda^\alpha \gamma_\mu\gamma_5 \psi. 
\end{equation} 
Combined with the vector fields
\begin{equation}
V_\mu^\alpha=\overline\psi \lambda^\alpha \gamma_\mu \psi,
\end{equation} 
we have the famous algebra of currents
\begin{eqnarray}
&&[V_0^\alpha(x),V_0^\beta(y)]=if^{\alpha\beta\gamma}V_0^\gamma(x)\delta(x-y)\cr
&&[V_0^\alpha(x),A_0^\beta(y)]=if^{\alpha\beta\gamma}A_0^\gamma(x)\delta(x-y)\cr
&&[A_0^\alpha(x),A_0^\beta(y)]=if^{\alpha\beta\gamma}V_0^\gamma(x)\delta(x-y)
\end{eqnarray}
with the $f^{\alpha\beta\gamma}$ being the structure constants for the
internal symmetry group.  

\endcomment
  
\section{Topology and chiral symmetry}

So far we have concentrated on aspects of the quantum theory.  But
non-perturbative phenomena also play a role in the classical
theory \cite{Yang:1954ek}.  With non-Abelian gluon fields,
a boundary condition such as the vanishing of the field $F_{\mu\nu}$
at spatial infinity does not require the gauge potential
$A_\mu\rightarrow 0$, but only that the potential be a gauge
transformation of zero field, i.e.
\begin{equation}
 A_\mu \longrightarrow -{1\over g^2}h^\dagger\partial_\mu h 
\end{equation}
where $h$ is a group valued function of space.  For our four
dimensional world, the space at infinity is sphere $S_3$ and the
function $h(x)$ can wrap non-trivially about this sphere.  The space
of smooth gauge fields separates into sectors labeled by this winding,
which can be determined directly from the gauge fields
\begin{equation}
{g^2\over 16\pi^2}
\int d^4x {\rm Tr}F\tilde F=\nu
\label{winding}
\end{equation}
where the index $\nu$ is an integer.  The path integral should sum
over all such sectors.  Of course, as mentioned earlier, the typical
gauge configuration is not smooth; indeed, it is not differentiable in
the continuum limit.  This introduces a small ambiguity in defining
the winding; we will return to this briefly in Section \ref{masses}.

These topological issues become particularly relevant upon introducing
fermion fields, i.e. the quarks.  Whenever $\nu\ne 0$ the Dirac
operator has zero modes satisfying
\begin{equation}
  D\psi(x)=\gamma_\mu (\partial_\mu+igA_\mu) \psi(x)=0.
  \end{equation}
Furthermore these modes are chiral
\begin{equation}
  \gamma_5 \psi(x)=\pm \psi(x).
\end{equation}
The basic index theorem relates the number of such modes
to the index in Eq. (\ref{winding})
\begin{equation}
  \nu = n_+-n_-
\end{equation}
where $n_+$ and $n_-$ are the number of positive or negative chiral
zero modes, respectively.

The importance of the zero modes was nicely interpreted in
\cite{Fujikawa:1979ay}.  Configurations with non-trivial winding exist
and must be included in the path integral.  On these formally
\begin{equation}
  {\rm Tr}\gamma_5 \equiv
  \sum_i \langle\psi_i \gamma_5 \psi_i \rangle=\nu\ne 0.
\end{equation}
where the sum is over a complete set of modes of the Dirac operator.
All other than the zero modes occur either in chiral-conjugate pairs
or ``above the cutoff'' and don't contribute to the sum.  On such
configurations the naive chiral transformation
\begin{equation}
  \psi \rightarrow e^{i\gamma_5 \theta}\psi
  \label{chiral}
\end{equation}
is not a symmetry.  Instead it changes the fermion measure in the path integral
\begin{equation}
 d\psi \rightarrow |e^{i\gamma_5\Theta}|\ d\psi
 =e^{i\nu\Theta}\ d\psi.
 \end{equation}

If we try to make the change of variables in Eq. (\ref{chiral}),
topology inserts a factor of $e^{i\nu\Theta}$ into the path integral.
This gives rise to a new and inequivalent theory.  This breaking of
the naive chiral symmetry is at the heart of why the $\eta^\prime$
particle is heavier than the pseudo-Goldstone mesons and its mass is
not directly proportional to the quark masses.  Indeed, $\Theta$ is a
hidden non-perturbative parameter for QCD.  For each value of
$\Theta$, the perturbative expansion is identical!  Perturbation
theory alone does not fully define the theory.

Note that the special case $\Theta=\pi$ reverses the sign of the
quark mass term
\begin{equation}
  m\overline\psi \psi \rightarrow -m\overline\psi \psi.
\end{equation}
This means that three flavor QCD with negative masses is a different
theory, i.e QCD at $\Theta=\pi$.  This is in strong contrast to
perturbation theory, wherein the sign of a fermion mass is merely a
convention.

The possibility of $\Theta\ne 0$ raises an as yet unresolved puzzle.
In this case CP symmetry is violated.  As no CP violation has been
observed in the strong interactions, experimentally $\Theta$ must be
very small.  Since CP violation is seen in the weak interactions, why
are the strong interactions different?

\section{Confinement and quark masses}
\label{masses}

Quarks are confined in hadrons.  What does their mass mean?  For a
physical particle, such as the proton, the mass follows from how it
travels over long distances.  We measure a particle mass seeing how its
energy and velocity are related
\begin{equation}
  E=mc^2+{1\over 2}mv^2+O(mv^4/c^2). 
\end{equation}
We can't use this relation for quarks since they don't propagate
alone.  Indeed the quark propagator is a gauge and scheme dependent
quantity.

The renormalization group integration constant $M$ would seem to be 
a natural candidate for a renormalized quark mass
\begin{equation}
  m=M\ g^{\gamma_0/\beta_0}(1+O(g^2))+\hbox{non-perturbative}.
\end{equation}
But some time ago 't Hooft \cite{'tHooft:1976fv} showed that the
``non-perturbative'' effects will mix the various $M_i$ and $\Lambda$
in non-trivial ways.  With a lattice regulator, this mixing can depend
on the details of the lattice regularization scheme.

Chiral symmetry provides one handle on this.  With degenerate light
quarks, we have light pseudoscalar bound states.
In particular, massless quarks imply massless pions.  For degenerate
quarks, the concept of a vanishing quark mass, $m_q=0$, is well
defined.

When the quarks are not degenerate, the situation is more complicated.
Simple chiral perturbation theory indicates that the mass of a
pseudoscalar is proportional to the average of its constituent masses,
as in Eq. (\ref{averageqm}).  Combining such relations with different particles
gives estimates for quark mass ratios. For example
\begin{equation} 
{m_u\over
m_d}\sim { 2m_{\pi^0}^2-m_{\pi^+}^2+m_{K_+}^2-m_{K_0}^2\over
m_{\pi^+}^2-m_{K_+}^2+m_{K_0}^2}\sim 0.55
\label{updown}
\end{equation}
suggests that the up quark has about half the down quark mass.  This
particular combination is chosen to reduce electromagnetic corrections
\cite{Dashen:1969eg}.

Eq. (\ref{updown}) must be understood in the context of several
caveats.  First, different meson combinations give slightly different
values for this ratio due to the chiral expansion being only an
approximation and higher order effects being ignored
\cite{Kaplan:1986ru}.  Second, when the quarks are not degenerate,
mixing occurs between the various neutral pseudoscalar mesons.  When
the strange quark is included, near the chiral limit the mixing is
primarily with the eta.  This is incorporated into the generalization
of Eq. (\ref{averageqm})
\begin{equation}
 m_{\pi_0}^2 \propto\  {1\over 3} \bigg(m_u+m_d+m_s
 -\sqrt{m_u^2+m_d^2+m_s^2-m_um_d-m_um_s-m_dm_s}\bigg).
 \label{etamixing}
\end{equation}
This is not the whole story since further mixing occurs with the eta
prime and pure gluonic states, particles that obtain the bulk of their
mass independent of chiral symmetry.  Concentrating on the lightest
two flavors, isospin breaking starts out quadratic in up-down mass
difference
\begin{equation}
  {M_{\pi_0} \over M_{\pi_+}} =
  1-O\left({(m_d-m_u)^2\over m_s^2},
{(m_d-m_u)^2\over \Lambda^2}
    \right).
\label{isobreaking}
\end{equation}
The low masses of the up and down quarks makes this splitting quite
small for the physical pions, even though the up quark is only about
half as heavy as the down.  The scale for the correction involves the
masses of the higher states.  For the mixing with the eta the
correction is contained in Eq. (\ref {etamixing}).  For the eta prime
and the glue states that scale is not determined directly by the quark
masses but involves the scale $\Lambda$.  One consequence is that
holding quark mass ratios fixed during renormalization is only
approximately equivalent to holding physical particle mass ratios
constant.

\begin{figure}[h] \em
\epsfxsize .5\hsize
\centerline{\epsffile{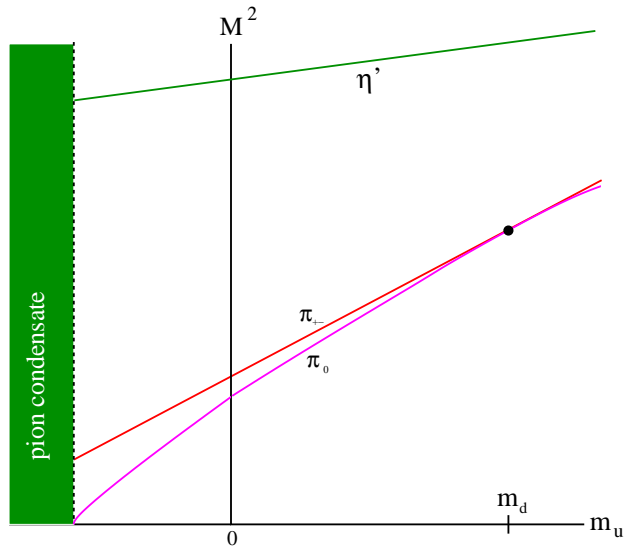}}
\caption{\label{iso2} As the up quark mass decreases to negative
  values, the neutral pion mass can pass through zero, giving rise to
  a $\pi_0$ condensate.}
\end{figure}

Although isospin breaking is small in practice, it is interesting to
consider what could happen if the light quark mass difference were
larger.  Note in particular that Eq.~(\ref{averageqm}) indicates that
a mass gap persists if only one quark, say the up quark, is massless
when the other quarks have positive mass.  Furthermore it appears that
physics remains sensible even if the up quark has small negative mass.
Remember, as mentioned earlier, in perturbation theory the sign of a
fermion mass is a convention.  This is not the case for QCD.

A particularly interesting situation occurs if the up quark mass is
negative enough, i.e. $m_u=O(-m_d)$.  The neutral pion decreases in
mass and eventually can pass through zero.  This gives rise to a
condensation of the $\pi_0$ field as sketched in Fig. \ref{iso2}.
This CP violating condensate is referred to as the Dashen phase
\cite{Dashen:1970et}.

\begin{figure}[h] \em
\epsfxsize .35\hsize \centerline{\epsffile{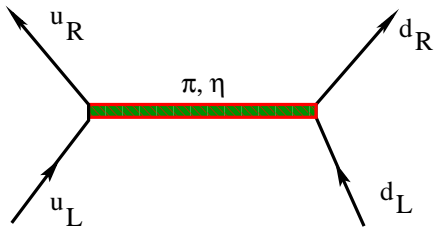}}
\caption{\label{fourpoint} A spin flip process involving both the up
  and down quarks occurs on exchanging pseudoscalar mesons.  The $\pi$
  and $\eta$ contributions cannot cancel since they are non-degenerate.}
\end{figure}

The neutral pion also provides a simple example of how confinement
entangles the quark masses.  This particle is a bound state of left
and right handed quarks and involves both up and down quark species.
If we create the $\pi_0$ from up quarks, $i\overline u_L \gamma_5
u_R$, and then destroy it using down quarks, $i\overline d_L \gamma_5
d_R$, we have a spin flip process, as shown in Fig. \ref{fourpoint}.
Unlike in perturbation theory, this is not suppressed at small light-quark
masses.  It is due to topology through what is called the ``'t Hooft
vertex'' \cite{'tHooft:1976fv}.

\begin{figure}[h] \em
\epsfxsize .35\hsize
\centerline{\epsffile{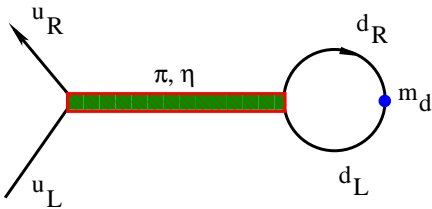}}
\caption{\label{induced2} A down quark mass can induce an effective
  mass for the up quark.}
\end{figure}

We see that the naive result that left and right handed quark fields
decouple at $m=0$ fails non-perturbatively.  Furthermore, this induces
a mass mixing for different quark species.  If we turn on a small down
quark mass in Fig. \ref{fourpoint} and close the down quark lines in a
loop, we arrive at Fig. \ref{induced2}.  This shows that a
non-vanishing down quark mass can generate an effective mass for the
up quark, even if it started off massless.  Indeed, topological
effects entangle all the quark masses and $\Lambda$ in a non-trivial
way.\footnote {Non-perturbative regulators for fermion fields always
  require unphysical parameters to control the chiral anomalies and
  what is known as the ``doubling problem.''  This brings an
  unavoidable scheme dependence.  With Wilson fermeions
  \cite{Wilson:1975id} there is the parameter ``$r$'' on which quark
  mass ratios can depend.  Overlap fermions \cite {Neuberger:1997fp}
  depend on a scheme dependent kernel.}

\begin{figure} \em
\epsfxsize .4\hsize
\centerline{\epsffile{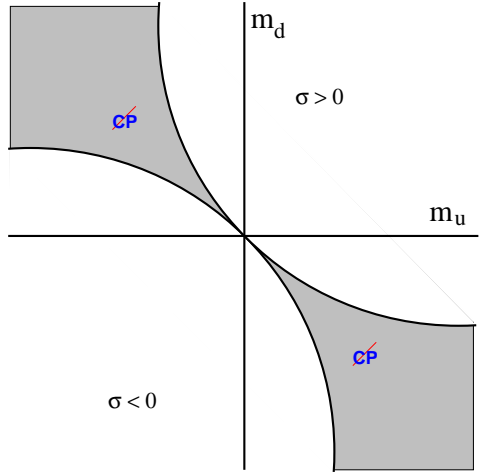}}
\caption{\label{iso4} The qualitative phase diagram of two flavor QCD
  as a function of the quark masses. Here $\sigma$ represents to the
  quark condensate $\langle\overline\psi \psi\rangle$.}
\end{figure}

Proceeding further through chiral perturbation theory, one can explore
the qualitative phase diagram shown in Fig. \ref{iso4} for the two
flavor theory as a function of the up and down quark masses, including
the CP violating Dashen phase \cite{Creutz:2010ts}.  This figure has
some interesting symmetries.  First of all, it is symmetric on
exchanging the up and down quarks.  This is related to isospin and
protects the quark mass difference $m_d-m_u$ from additive
renormalization; i.e if the quarks are degenerate, they remain so
through the regularization process.  It is also symmetric under
flipping the sign of both quark masses, $(m_u,m_d)\rightarrow (-m_u,
-m_d)$.  This is a consequence of the flavored chiral symmetry
$\psi\rightarrow e^{i\pi \tau_3\gamma_5}\psi$, a good symmetry since
$\tau_3$ is traceless.  This in turn protects the average quark mass
from an additive renormalization.

It is important, however, to note the absence of any symmetry on
flipping the sign of a single quark mass, for example
$(m_u,m_d)\rightarrow (-m_u,m_d)$.  Away from $m_d=0$ there is no
singularity at $m_u=0$.  This is a strictly non-perturbative effect
and shows that a non-degenerate massless quark is not protected from
renormalization.
%In particular, the possibility of $m_u=0$ is
%unrelated to the strong CP problem as long as $m_d\ne 0$.
Should we care if the concept of a single vanishing quark mass is a
bit fuzzy?  This quantity is not directly measured in any scattering
process.  This issue is closely tied to the ambiguities in defining
gauge field topology since typical gauge fields are non-differentiable
% \cite{Creutz:2013xfa}.
Only real particle masses and their
scatterings are physical.

\section{Conclusion}

Going beyond perturbation theory is crucial for understanding QCD.
Lattice gauge theory augmented with ideas from renormalization merge
to give a coherent picture accommodating non-perturbative concepts
such as confinement and chiral symmetry breaking.  Variations of QCD
with identical perturbative expansions can display different physics.
This includes the situation of a non-vanishing Theta as well as the
peculiar possibility of negative quark masses.  Finally, the numerical
values of the quark masses, as well as their ratios, can depend
subtlely on the details of the regularization scheme.

%-----------------------------------

%\appendix
%\section{}

%\begin{acknowledgments}
%\end{acknowledgments}


%\bibliography{/home/mike/papers/references,/home/mike/papers/creutz}
% used bibexport to generate references.bib
\bibliography{references}

\end{document}